\documentclass{jpp}
\usepackage{graphicx}
\usepackage{epstopdf, epsfig}

\usepackage[utf8]{inputenc}
\usepackage[T1]{fontenc}
\usepackage{amsmath}

\shorttitle{Exploring zonal flow mediated saturation on stellarators}
\shortauthor{C.D. Mora Moreno, J.H.E. Proll, G.G. Plunk, P. Xanthopoulos}

\title{Exploring the dependence of zonal flow mediated saturation on the magnetic geometry of stellarators}

\author{C. D. Mora Moreno\aff{1}
  \corresp{\email{c.mora.moreno@tue.nl}},
  J.H.E. Proll\aff{1},
  G.G. Plunk\aff{2},
 \and P. Xanthopoulos\aff{2}}

\affiliation{\aff{1}Eindhoven University of Technology, P.O. Box 513, 5600 MB Eindhoven, The Netherlands
\aff{2}Max-Planck-Institut für Plasmaphysik, Wendelsteinstrasse 1, 17491 Greifswald, Germany}

\begin{document}

\maketitle

\begin{abstract}
In stellarators, zonal flow activity depends sensitively on geometry of the three dimensional magnetic field, via an interplay of mechanisms that is not fully understood. In this work, we investigate this by studying three magnetic configurations of the Wendelstein 7-X stellarator.  We find that variation in linear zonal flow damping is accompanied by variation in nonlinear drive, and identify key geometric features that control these effects. Understanding the resulting balance is important for the development of reduced models of turbulent transport.
\end{abstract}

\section{Introduction.}

In magnetically confined plasmas, turbulence is responsible for driving unwanted transport of particles and energy. This has an impact on the delicate confinement in the core region, needed for optimal fusion power yield. Turbulent fluctuations, of the scale of the ion gyro-radius, develop due to linear instabilities driven by the free energy available in strong gradients of temperature and density. These instabilities are key to understanding the nonlinear character of the turbulence.

In stellarators, linear instabilities are particularly influenced by the structure of the magnetic field, making this a decisive feature of the device. The stellarator concept has significant advantages to tokamaks. With their distinctive 3-dimensional magnetic field and higher amount of degrees of freedom, stellarator configurations present the opportunity to find the configuration with best confinement properties. As an example, the magnetic field geometry (from now on "field geometry") has been successfully optimised for minimal neoclassical transport \citep{Grieger1992}, resulting in the most advanced class of these devices.

Stellarators may in the future be further optimised for lower turbulent transport. The increased flexibility of the field geometry becomes both an advantage and an obstacle for this optimisation. A clear perspective is needed, of the physical mechanisms involved in turbulence saturation. This can serve as a basis for developing reduced models to help explain and predict the transport observed in nonlinear turbulence simulations.

In this work we focus on turbulence caused by the ion-temperature-gradient (ITG) mode, one of the most common turbulent modes and the cause of large transport losses in fusion plasmas \citep{Cowley1991}. Building from previous work on ITG turbulence in stellarators \citep{Plunk2014a,Plunk2015b, Plunk2017}, we explore the dynamics of ITG saturation, employing lessons from linear physics for understanding the observed transport.

Zonal Flows (ZFs) arise as secondary instabilities from ITG turbulence. These secondary structures are elongated in the poloidal direction and have the ability to shear, and suppress the growth of turbulent vortexes. Since ZFs feed on the energy of the driving instabilities, they are an inherently nonlinear phenomenon. The ZF mechanism acts as a small-scale transport barrier, resulting in increased plasma confinement. This mechanism is of utmost importance for turbulence saturation in tokamaks.

A natural starting point for the study of ZFs is analysis of the linear response of the component of the electrostatic potential. This linear response is characterised by the decay of the ZF component of the electrostatic potential ($k_{y}=0$, $k_{z}=0$, where $k$ is the wave-number in the bi-normal and parallel directions respectively). The analytical expressions derived by \citet{Rosenbluth1998} aimed to predict the final state of the linear ZF response, called residual. In their work, they related this residual to large scale quantities, such as aspect ratio and safety factor.

While ZFs have been studied linearly in stellarators, their nonlinear interaction with the turbulent mode is not yet fully understood, and certainly not at the level required for predictive transport modelling. Some important results about the linear response nevertheless give a basis for this understanding. As a generalisation of the work by \citet{Rosenbluth1998}, \citet{Sugama2005} proposed an analytical derivation of the response kernels to predict the decay and residual of the linear ZF response.

Analytical results obtained by \citet{Gao2008a} predict that the decay of the linear ZF response ($\gamma$) is proportional to the parallel wave-number ($k_{\parallel}$). This relation describes the resonance of the sound wave component in the geometry itself, forming an oscillation that decays in time. \citet{Mishchenko2008a} identified an additional mechanism, unique to stellarators, giving rise to solutions with slow algebraic decay and a characteristic real frequency. The residual of the linear ZF response has been linked to the level of ZF activity in some predictive models. These models, while successful in tokamaks, fail to be predictive if the residual is close to zero.

In the stellarator context, \citet{Nunami2013} demonstrated that a decay rate extracted from linear zonal flow response can be a good predictor of ZF activity in the LHD stellarator. However, with the multiple distinct decay processes, acting on different timescales, the evaluation of different stellarator geometries is needed to determine which decay rate, if any, can be a general predictor of ZF activity. It may be argued, in cases where the slow decay time is much smaller than nonlinear correlation times, that the initial response would be the more relevant measure. A more complete understanding of ZF physics will certainly allow for better transport prediction.

\begin{figure}
    \centering
    \includegraphics[width=0.80\textwidth]{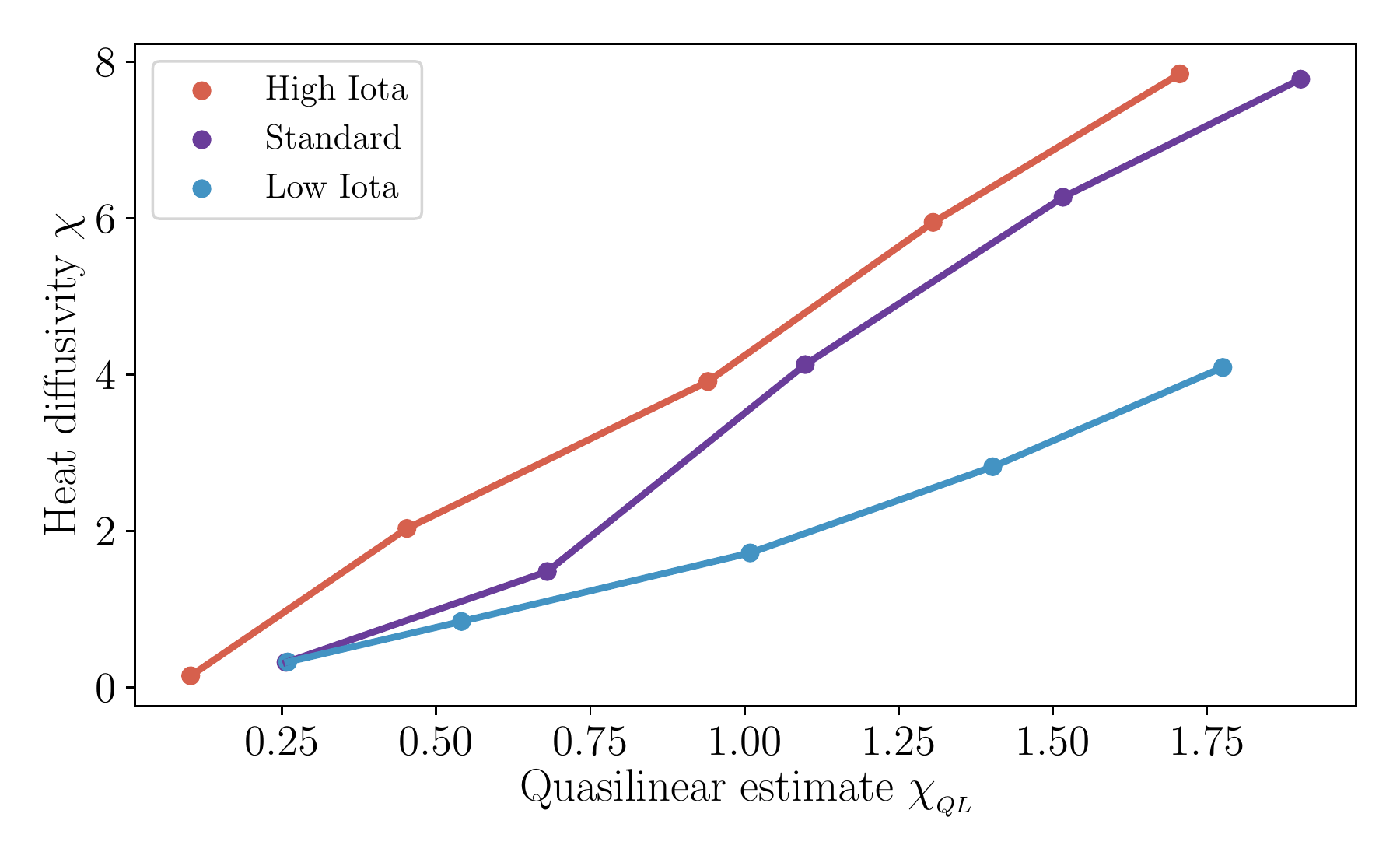}
    \caption{A simple quasilinear estimate provides an incomplete answer for the heat diffusivity prediction and will be configuration-dependent. Here we observe three configurations of W7-X, in which the pre-factors such as slope and scaling factors are missing. See the text for the formulation of the quasilinear estimate.}
\label{fig:ql_estimate}
\end{figure}

A typical approach to transport prediction is quasilinear theory, in which the transport estimates are derived solely from linear calculations. As shown in figure \ref{fig:ql_estimate}, such modelling is incomplete. There is certainly a proportional relation between the heat diffusivity ($\chi$) and the quasilinear estimate ($\chi_{_{QL}} \approx \sum_{k_{y}} \gamma_{_{k_{y}}} / k^{2}_{y} \times \Delta k_{y}$, where $\gamma_{_{k_{y}}}$ is the linear mode growth rate), but the pre-factors that determine the slope are related to the field-geometry-dependent mechanisms responsible for turbulence saturation.

Recapitulating, the saturation mechanisms must be understood for the creation of better transport models with aims for turbulence optimisation. Progress in this direction is exemplified by the notable proxies derived by  \citet{wendel} and \citet{Mynick2014a} which were successful, but limited to linear predictions. Regarding the development of advanced proxies, they mention that higher effectiveness will be given by the identification of the key geometric components that control turbulence saturation. The effectiveness of future proxies and methods will be higher if they are developed from a physics-based understanding, rather than a brute-force approach.

The basis of improved transport modelling that we propose compiles the relations between nonlinear ZF formation, turbulence saturation, and linear predictions; as determined by the stellarator field geometry. We are thus motivated by the encouraging results on the LHD stellarator, by \citet{Nunami2012, Nunami2013}, to generalise the relations observed between the linear and nonlinear dynamics of ZFs.

The ground for our investigations is formed by three different configurations of the Wendelstein 7-X (W7-X) experiment \citep{Klinger2013}, these are: High Iota, Standard, and Low Iota. Within these, we obtain the rotational transform (iota) as a macroscopic variable to understand how it affects the saturation dynamics. Alongside, we see the influence of two other geometric characteristics: The geodesic curvature ($\kappa_{G}$) and the drift well ($\mathcal{K}_{1}$), both defined by \citet{Xanthopoulos2009}. The latter quantity was reported by \citet{Plunk2014a}, to determine the localisation of the turbulent mode and affect the generation of ZFs.

The present work proceeds as follows. First, we describe the geometric characteristics and the quantities of interest. We then present the details of our numerical simulations and our methodology. The results section opens with the analysis of the linear ZF response, where the configurations with strong linear ZF damping are expected to have weak nonlinear ZF levels. We proceed to address the question ``How important do ZFs become when the rotational transform is increased or decreased?", by performing numerical experiments in which the ZF component is artificially suppressed. We find that the Low Iota configuration, with a short geodesic length and long connection length, is more sensitive to the absence of ZFs, contrary to the prediction from the linear ZF decay. This contradiction motivates the study of the toroidal extension of the turbulent mode, where we observe that peaked modes are unable to generate ZFs efficiently. Finally, we relate the characteristic scales from the field geometry to linear- and nonlinear ZF responses. The work concludes with the basis for improved stellarator transport prediction.

\section{Magnetic field characteristics.}

We provide here more detail about the configurations and components of the field geometry. Note that within the configurations presented, the Standard case serves as a reference. In the rest, as the names suggest, the radial iota profile of has either high or low values. But since our simulations are local, iota has a is a number in the same trend (high, medium and low). Iota describes the average poloidal transit of the field line per toroidal transit. As we show later, the "twist" is closely related to the characteristic arc length, or physical distance along the field line.

\begin{figure}
  \centering
  \includegraphics[width=0.80\textwidth]{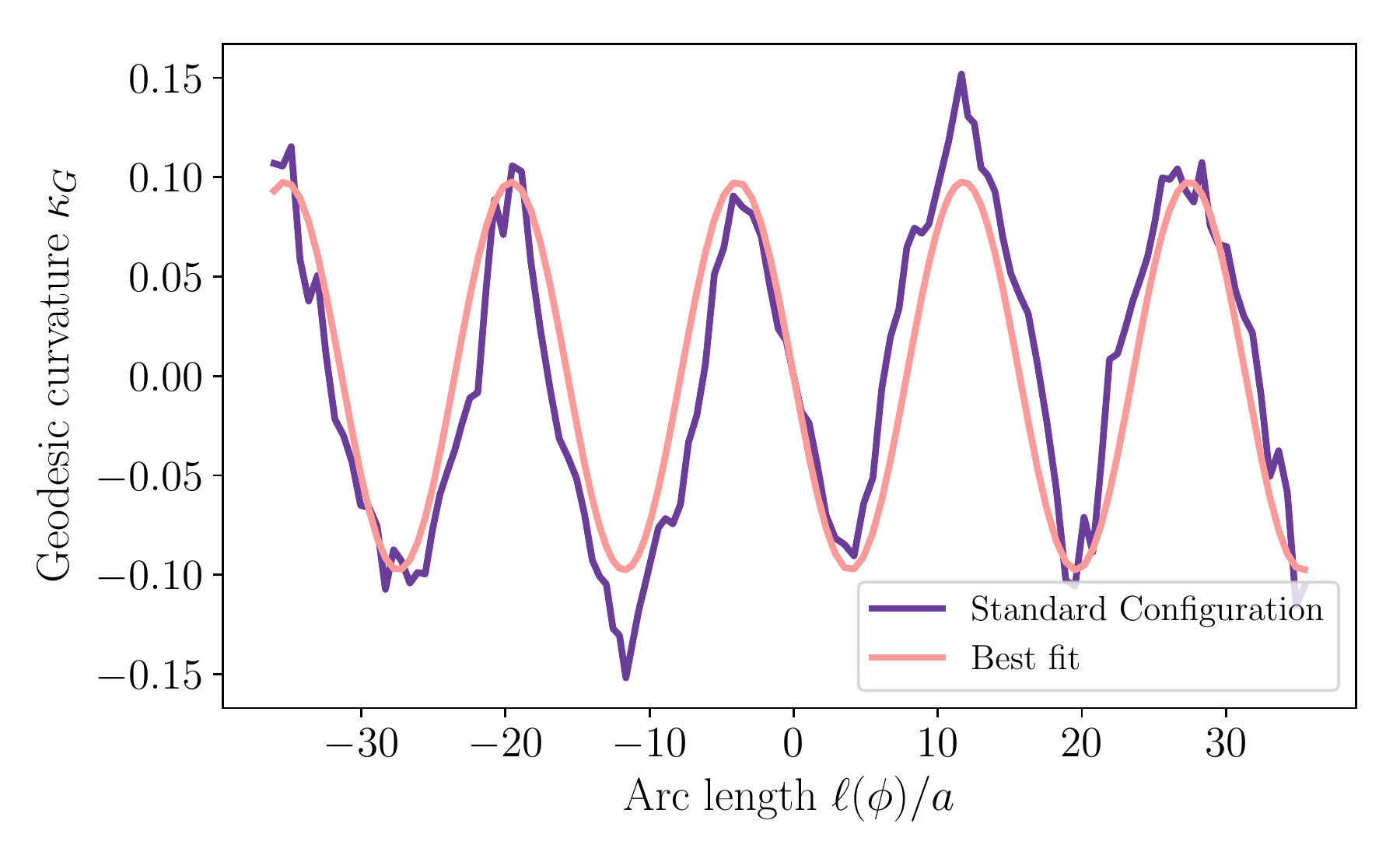}
  \caption{Geodesic curvature is the most relevant quantity for the linear ZF decay. Here we observe that of our reference case, the Standard Configuration. Tokamak theory predicts a characteristic length $L_{G}$ related to the periodicity. This quantity can be used to predict zonal flow behaviour, as explained in the text.}
  \label{fig:geodesic_curv}
\end{figure}

The first quantity of interest is the geodesic curvature, which enters the kinetic equation and couples the zonal component of the electrostatic potential to the sound wave. It is known \citep{Xanthopoulos2011a} that the magnitude of $\kappa_{G}$ is related to the decay of the linear ZF response. This geometric quantity in the Standard configuration is shown in figure \ref{fig:geodesic_curv}, plotted against the arc length ($\ell / a$) along the field line. Since flux tubes are generated within one poloidal turn (the short way around a torus), the high rotational transform results in a shorter arc length, or put differently, a shorter characteristic length scale over which the geodesic curvature varies.

\begin{figure}
  \centering
  \includegraphics[width=0.80\textwidth]{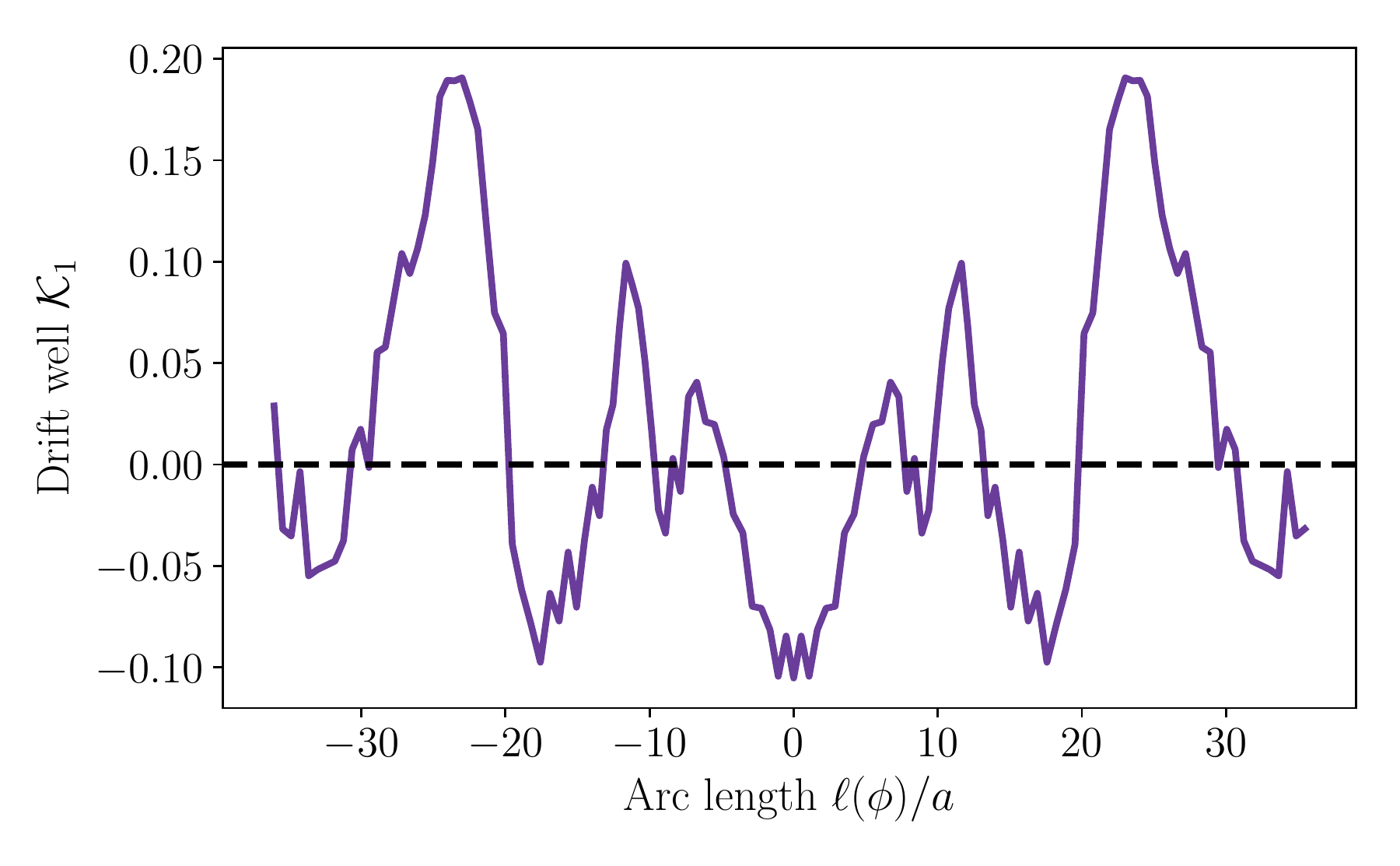}
  \caption{The drift well ($\mathcal{K}_{1}$) exists within the context of the gyrokinetic theory, and is defined in Eq. (147) by \citet{Xanthopoulos2009}. This quantity is related to the normal curvature, therefore called "bad" curvature due to its relation to the shape of the turbulent modes.}
  \label{fig:normal_curv.}
\end{figure}

We can approximate the geodesic curvature with the periodic form:
\begin{equation*}
  \kappa_{G, fit} = |A|\text{sin}(2 \pi \ell / L_{G})
\end{equation*}
\noindent where $|A|$ is the amplitude of the waveform and $L_{G}$ is a characteristic geodesic length. This allows us to use tokamak theory for a rough understanding, by noting that the wave-number of the sound wave, induced by geodesic coupling, can be estimated as $1/(q_{eff} {R_eff}) = k_{\parallel} = 2\pi/L_{G}$. The obtained characteristic length $L_{G}$ is unique for each configuration and flux tube, $L_{G}$ in the Standard configuration can be seen in figure \ref{fig:peakdist} as the arc length necessary for a complete oscillation ($\sim$ 15.7).

It has been proposed \citep{Plunk2015b} that the turbulent mode along the toroidal direction is spread in the valley of the "bad" or, normal curvature, forming a drift well (observed in their figure number 5). The implications of the broadening of the turbulent mode affect directly the generation of ZFs. If a mode is well spread over the toroidal direction we would expect a more efficient generation of ZFs and vice versa.

\section{Numerical Investigations.}

Plasma micro-turbulence in the core region of a magnetic confinement device is best described within the framework of gyrokinetic theory. The numerical simulations we present here were performed with the continuous code GENE \citep{Jenko2000c, Merz2008b, Xanthopoulos2009} using a flux tube domain \citep{Beer1995}. In the code a plasma distribution function is evolved in time, described by $g(k_{x}, k_{y}, z, v_{\parallel}, \mu, t)$, where $k_{x}$ and $k_{y}$ are the radial and bi-normal wave numbers respectively, $z$ is the field-aligned coordinate, $v_{\parallel}$ is the velocity in such direction, $\mu$ the magnetic momentum, and $t$ is time. Typical spatial resolutions of a flux tube in W7-X are (225, 96, 128, 32, 16), in the corresponding directions.

For the employed flux-tubes in the field geometry of W7-X, we chose a single radial position for all of them at the normalised radius $s = 0.5$, centred on the bean plane with field line label $\alpha = 0$ (we only note negligible changes with respect to the triangle plane).

The generation of ZFs and their interaction with the driving instability, as mentioned, is a nonlinear effect. We present three kinds of simulations in the present work.

The first, nonlinear simulations, are solutions of the full gyrokinetic system of equations. Transport quantities are taken from the saturated phase, with a sufficiently broad time window for significant statistics. From the heat transport $Q$, we obtain the heat diffusivity ($\chi$) by the following relation: $\chi = Q / n L_{T}$, where n is the density and $L_{T}$ the temperature gradient defined as: $L_{T}^{-1}=-\left(1 / T_{0}\right) d T_{0}/dr$. The results we present here belong in the far-from-marginal temperature gradients, where strongly-driven turbulence develops.

In the second kind of simulations we exclude the nonlinear terms of the gyrokinetic equation and excite all the modes in the presence of a temperature gradient. These purely linear simulations provide us with the growth rate of a singular driving mode, at particular values of the bi-normal wave-number $k_{y}$.

\begin{figure}
    \centering
    \includegraphics[width=0.80\textwidth]{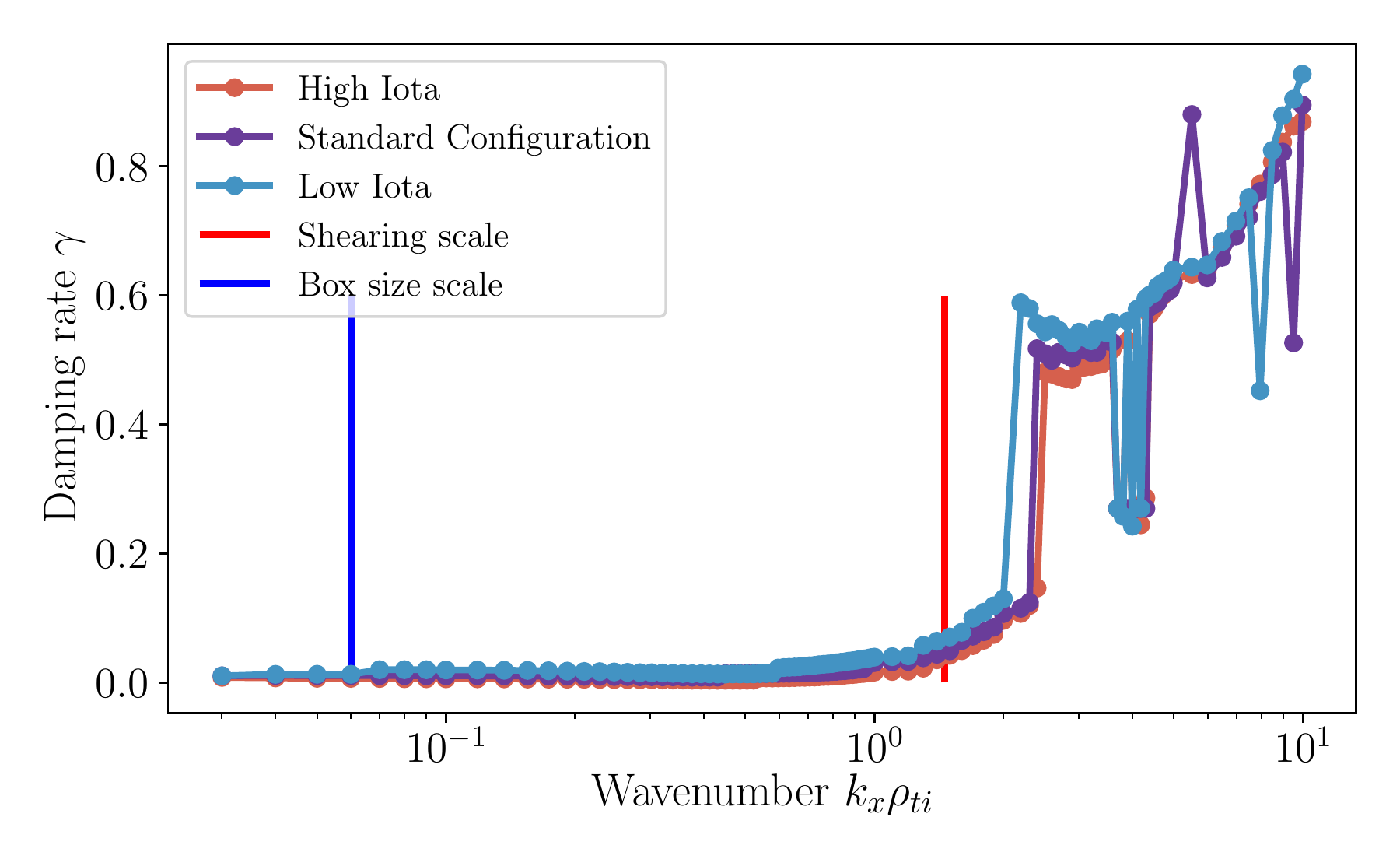}
    \caption{The damping rates of the linear ZF responses are affected by the selection of the wave-number. At the scales of the simulation domain (vertical blue line), the variations of the damping are minimal. Note that value of damping at the scale of maximal $\textbf{E} \times \textbf{B}$ shear is quantitatively different, but qualitatively similar.}
    \label{fig:kx_dep}
\end{figure}

The third kind of simulations correspond to the linear ZF response. When the nonlinear terms of the gyrokinetic equation are omitted –those related to the Reynolds stress– it is possible to study the linear evolution of the ZF component of the electrostatic potential, as a response to an initial perturbation. The resulting time-trace depends mainly on the size of the flux tube in the perpendicular direction $L_{x} = 2\pi/k_{x}$. The structures formed by ZFs have two scales, the flow profiles exist in the scale of the simulation domain, while the shearing of turbulent vortexes happens in the scale of the ion-Larmor-radius (see the vertical markers in figure \ref{fig:kx_dep}).

\begin{figure}
    \centering
    \includegraphics[width=0.80\textwidth]{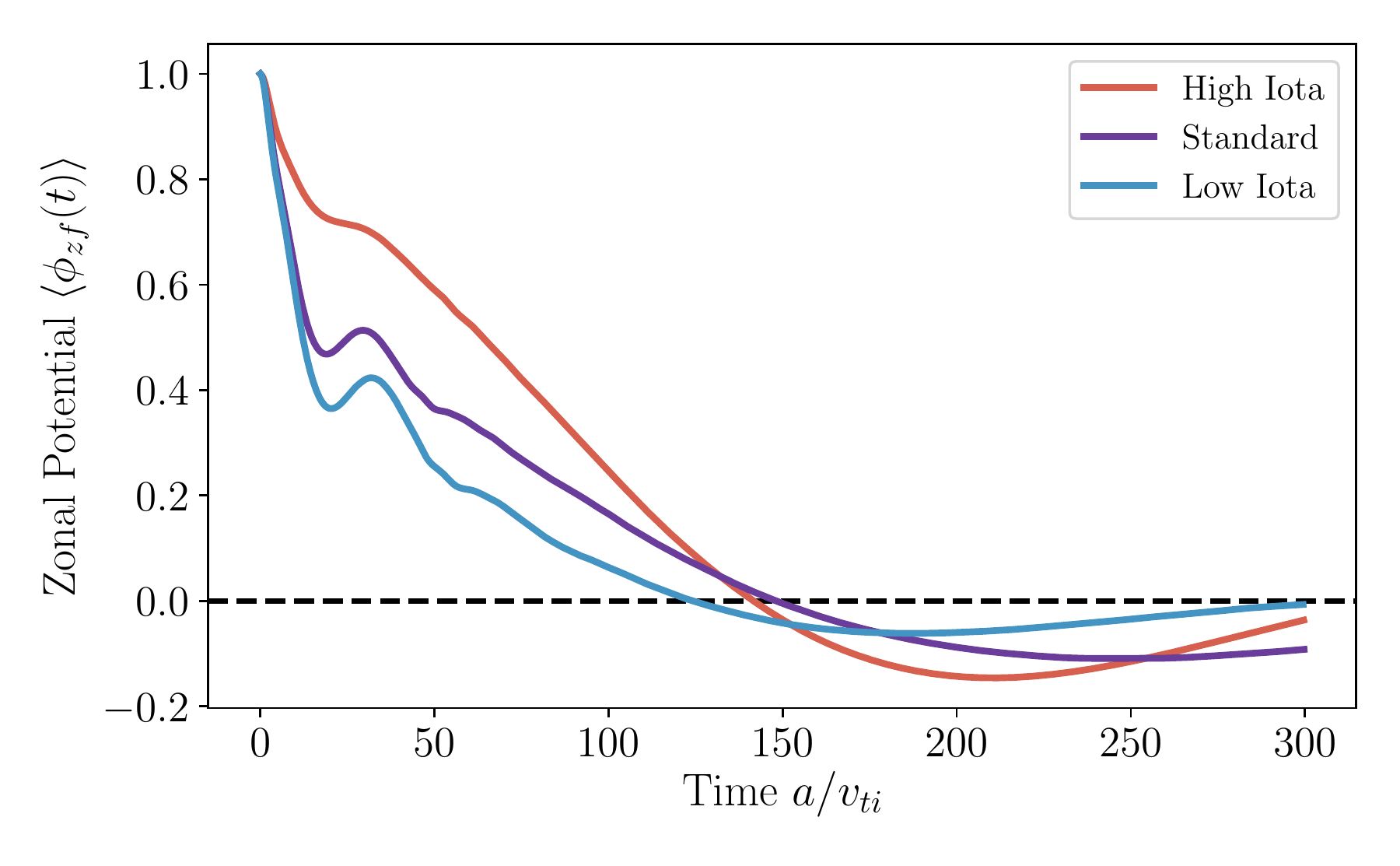}
    \caption{Early times of the linear ZF response in three different configurations of W7-X. The residual in all cases approaches zero as $Time \rightarrow \infty$. The $k_{x}$ wave-numbers were chosen here according to the size of the simulation domain. As seen in figure \ref{fig:kx_dep}, there is a small variation of the initial damping at large scales.}
    \label{fig:linear_ZF_response}
\end{figure}

As mentioned, we are interested in early times of the linear ZF response and not in the long-time solution. The fast dynamics in the early times of this response are relevant in nonlinear saturation. To capture this, an exponential curve with an oscillating part was fitted, following the equation:
\begin{equation*}
    \phi_{ZF, fit}(t) = (A - d * t) + ((1 - (A - d * t))  \text{exp}(-\gamma * t) * \text{cos}(\omega * t))
\end{equation*}
\noindent where $A$ defines the amplitude, $d$ is a linear decay, $\gamma$ is the exponential decay, $\omega$ is the oscillation frequency, and $t$, time. The curve-fitting process was carried out by a non-linear least-squares minimisation routine \citep{Newville2020}, with a time window from 0 to 6 time units. Increasing the upper time limit of the interval forces the fitting routine leads to ignore the fast dynamics we deem relevant.

\begin{figure}
    \centering
    \includegraphics[width=0.80\textwidth]{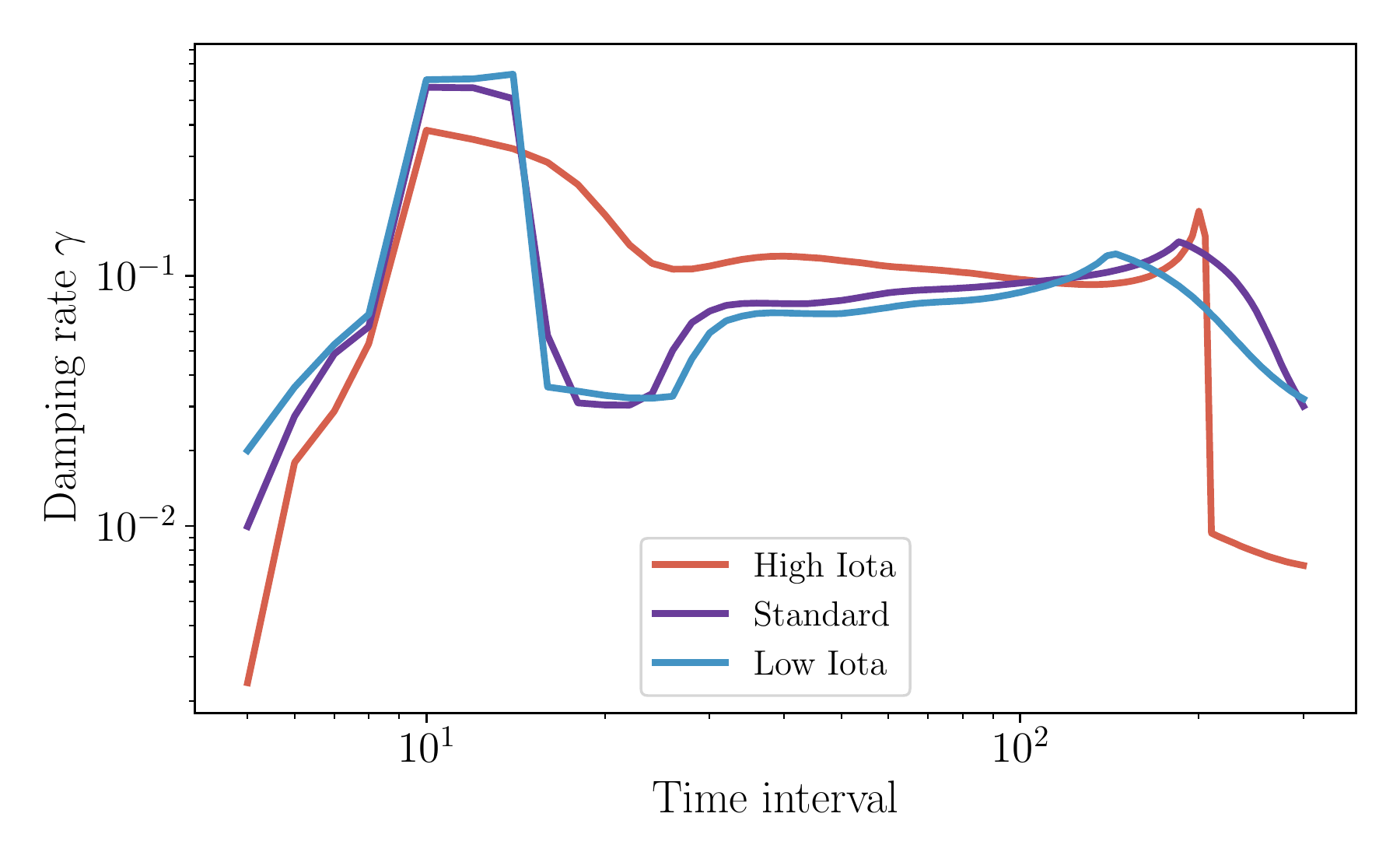}
    \caption{Linear ZF damping rate as a function of the upper limit of the fitting time interval. The time interval is central to capture the early potential drop. We used $t_{final} \sim 6$ in our results. A larger upper limit leads to an inverted order of the configurations, i.e. High Iota depicts a higher damping in the long time scales. Past 200 time units, the fitting routine is unable to describe the oscillation. Despite the apparent low relevance of the fitting process, fast vs slow dynamics of the linear ZF decay are described in this plot.}
    \label{fig:interval}
\end{figure}

\section{Results.}

\begin{figure}
    \centering
    \includegraphics[width=0.80\textwidth]{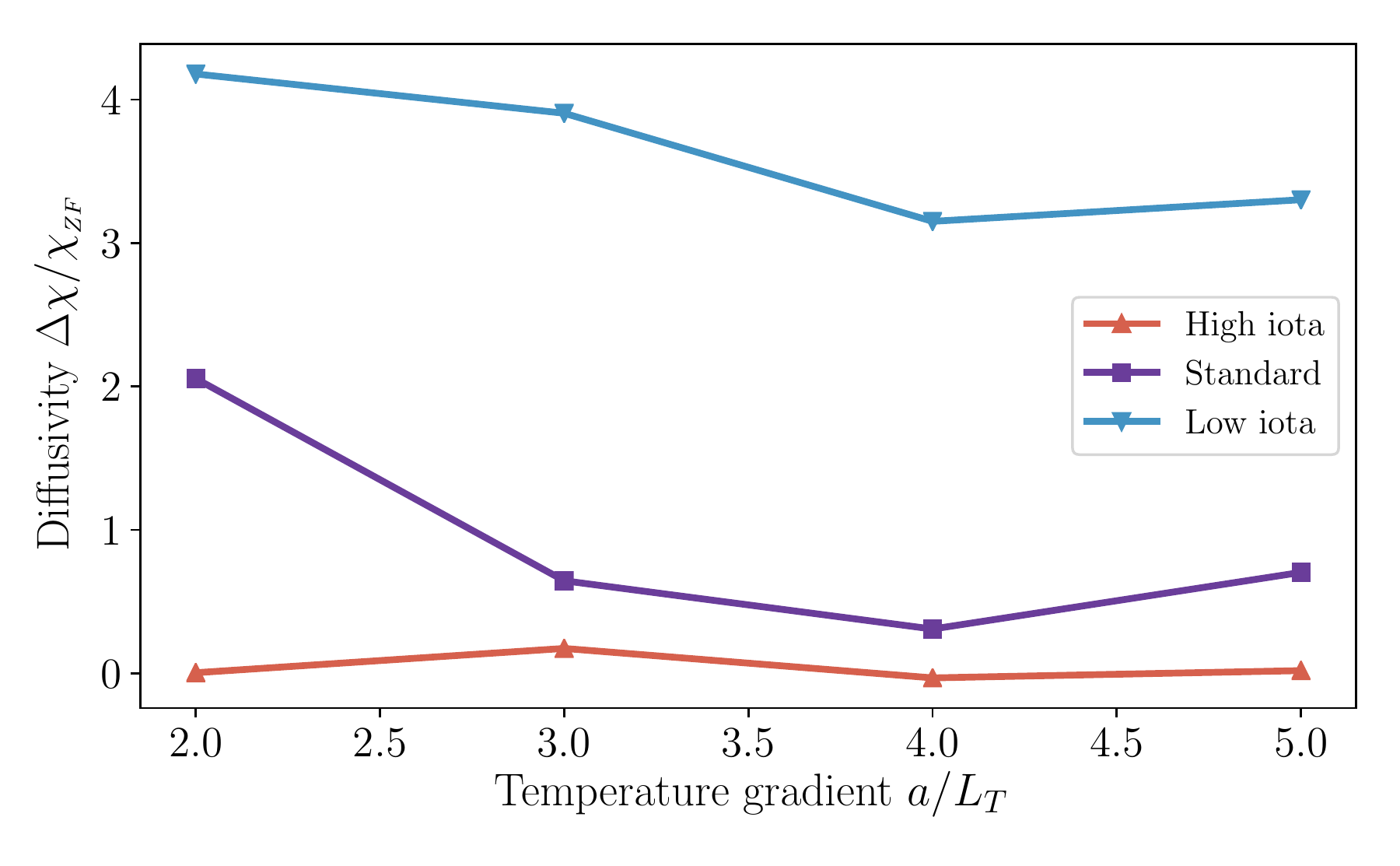}
    \caption{When the ZF component of the electrostatic potential is suppressed artificially in nonlinear simulations we observe how the configuration depends on it for saturation. This is observed as a change in diffusivity between simulations with and without ZFs for different gradients in the High Iota, Standard, and Low Iota configurations.}
    \label{fig:delta_chi}
\end{figure}

The presented results are valid in the strongly-driven turbulence regime, which develops at high temperature gradients, shown in figure \ref{fig:delta_chi}. Close to marginal stability we observe that a strong ZF regime develops and the geometric quantities we describe have no clear relationship to the mode shape.

The linear ZF response has been used as a proxy for transport prediction, both in tokamaks and other stellarators \citep{Nunami2013}; It has been observed in tokamak simulations that GAM damping and residual levels can influence the ZF activity, depending on the given parameter regime \citep{Waltz2008}. Note that the time trace yields a residual that is close to zero in all the configurations of W7-X.

\begin{figure}
    \centering
    \includegraphics[width=0.80\textwidth]{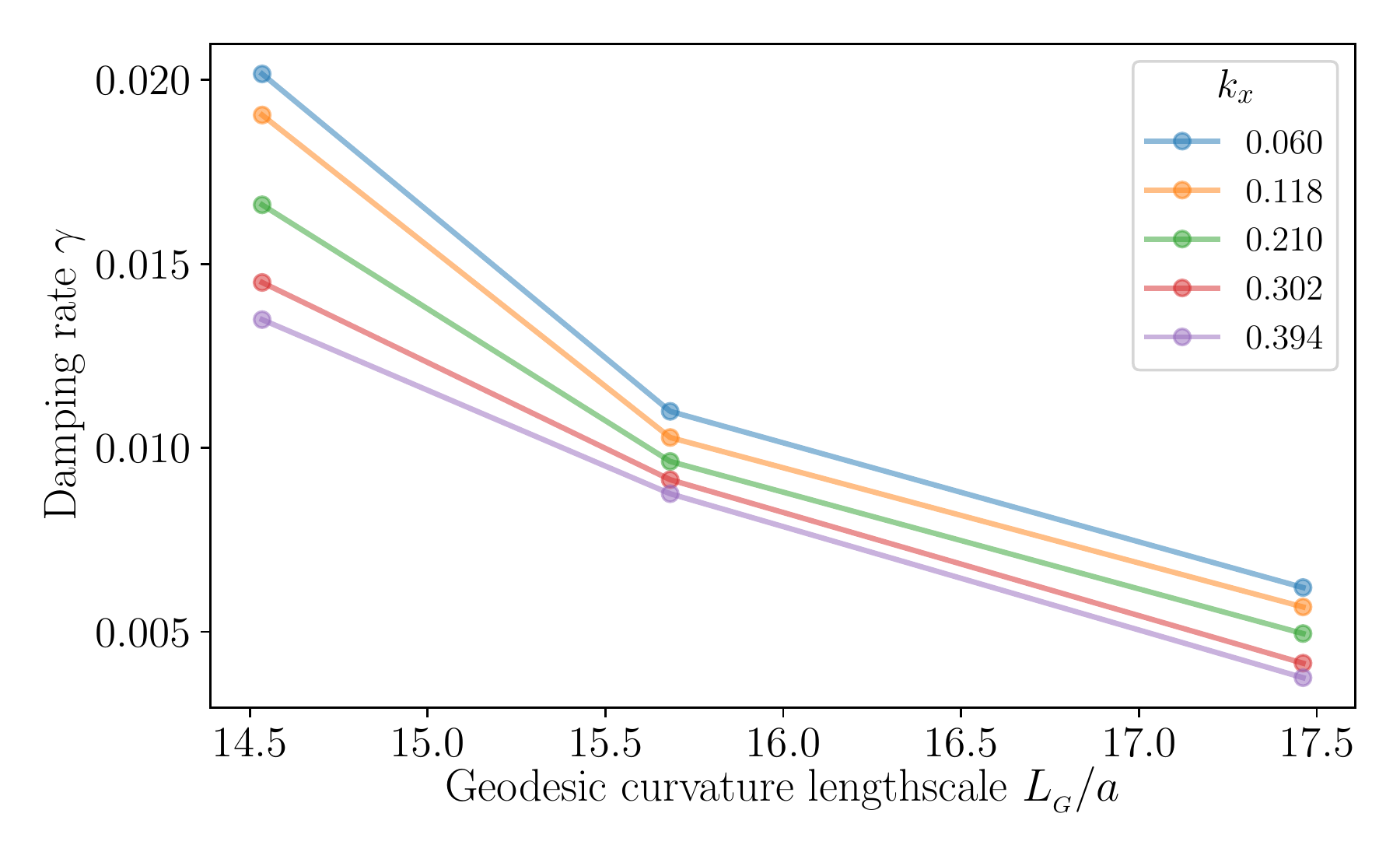}
    \caption{In the x axis, the increasing Geodesic curvature length-scale corresponds to Low Iota, Standard and High Iota configurations, where damping rate is higher, median, and lower respectively. With higher damping, a weaker nonlinear ZF activity is then expected, particularly in the Low Iota configuration.This relation holds for large scales of the size of the simulation domain and agrees with theoretical predictions \citep{Gao2008a}}
    \label{fig:peakdist}
\end{figure}

In figure \ref{fig:linear_ZF_response} we can observe the early times of the linear ZF potential decay. Since the linear response also depends on the length of the computational domain in the bi-perpendicular direction $k_{x}$, we choose to study large scale $L_{x}$ values (or, small $k_{x}$), given the macroscopic nature of the flows. In figure \ref{fig:interval} we see how the time interval determines the ZF dynamics characterised by the decay constant. As derived by \citet{Gao2008a}, the initial damping rates of the linear ZF response scale as $k_{\parallel} = 1 / L_{G}$ (seen in figure \ref{fig:peakdist}). These early times contain the dynamics of interest, related to the short timescale of turbulence saturation.

The transient damping rate is obtained from an exponential fit on the early decay of the linear ZF response. This quantity is found to be inversely proportional to the characteristic geodesic length, seen in figure \ref{fig:peakdist}. Even though the entire linear ZF response is determined by the field geometry, and this simultaneously sets the nonlinear transport, there is no direct physical reason to expect that either, the early decay or residual alone, should determine the nonlinear ZF efficiency. This relation suggests that configurations with lower geodesic curvature lengths will present lower ZF activity, since the damping is higher. In contrast with the geodesic curvature, we observed that artificially increasing or decreasing the flux surface compression $g_{11}$, related to ZF decay, has no direct impact on the linear ZF potential decay nor residual.

Having established that the geodesic curvature has – at least linearly – a significant effect, we proceed to investigate the influence of the ZF component for nonlinear transport reduction. We used the feature of the GENE code to set the ZF component of the electrostatic potential to zero every time step. The result is a change in heat diffusivity ($\Delta \chi$), normalised to the saturation value obtained in the presence of ZFs ($\chi_{_{ZF}}$). We report this in figure \ref{fig:delta_chi}. The change in diffusivity is higher in the Low Iota case, thus ZFs must to play a fundamental role for saturation, whereas in the High Iota case the diffusivity is not significantly changed. This behaviour suggests that ZFs play a negligible role in saturation.

Our results up to this point suggest that, in configurations with low $L_{G}$ and iota, ZFs will play a minor role in transport suppression. At the same time, we know from theory that a lower $L_{G}$ has been linked to stronger linear ZF damping, which would make us expect weaker ZFs nonlinearly. But from figure \ref{fig:ql_estimate} we observe the opposite. To rule out another mechanism being at play, we investigate ZF generation by the primary instability, ITG.

\begin{figure}
    \centering
    \includegraphics[width=0.80\textwidth]{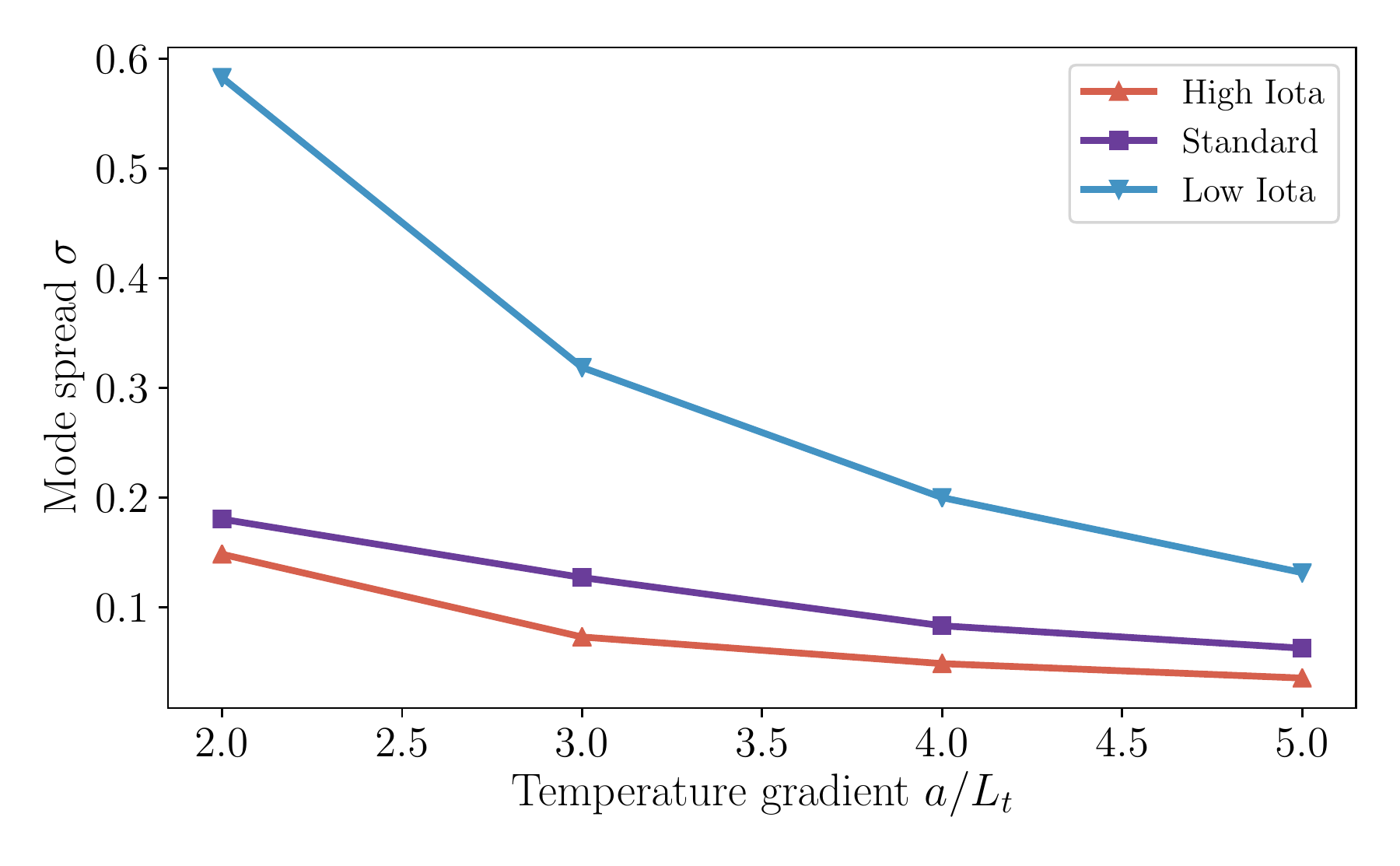}
    \caption{The mode spread ($\sigma$) represents the extension of the turbulent mode along the toroidal direction. An evenly spread mode, with high $\sigma$, will generate ZFs over a larger part of the flux tube.}
    \label{fig:mode_spread}
\end{figure}

Following the the predictions made by \citet{Plunk2015b, Plunk2017} regarding mode localisation, we investigated how the shape of the turbulent mode along the toroidal direction is related to our results. The $\sigma$ quantity lets us evaluate how spread out the turbulent mode is. It can be thought of as a filling factor, that measures the fraction of the turbulence domain that is occupied by the mode. If the mode is spread along the toroidal direction, ZFs might be generated more efficiently. The expression for $\sigma$ is found in equation (2) of \citet{Plunk2017}. In our results we observe strong variation of $\sigma$ across configurations, which is correlated with the depth and width of the drift well, finding an agreement with \citet{Plunk2014a}.

In figure \ref{fig:mode_spread} we observe a large spread of the mode in the Low Iota configuration and a smaller spread in the High Iota configuration. Both the spread and amplitude of the turbulent mode are closely related to the shape of the drift well. This relationship, as well as the quantitative relationship between geodesic curvature and linear ZF damping, will be explored in a future publication.

Peaked turbulent modes (with low $\sigma$) arise in particularly deep and narrow drift wells. Consequently, the strong localisation is unfavourable for the uniform generation of ZFs. Thus, we can make a direct link between efficient energy transfer between the ITG mode and the drive of zonal flows through the drift well length and turbulent mode spread.

To summarise, we identify two key field geometry quantities that influence the behaviour of ZFs: Geodesic curvature (in particular its characteristic scale of variation $L_{G}$) and the drift well, which contains the "bad" curvature.

\section{Conclusions.}

We have presented a study of the ZF generation and decay among three configurations of the W7-X stellarator which can be summarised as follows. Although higher average geodesic curvature and short geodesic length are related to higher linear ZF response damping in the initial times, this feature alone fails to be directly predictive of the nonlinear ZF activity in stellarators. We propose that geodesic and normal curvature must be integrated to account for variation in both ZF generation and decay mechanisms.

In nonlinear simulations, we see that transport suppression by ZFs is higher when the characteristic geodesic length and rotational transform are low. In such configurations, lower transport and high ZF dependence are observed.

Evenly distributed mode structures (high $\sigma$) in the toroidal direction are found when the shape of the drift wells is deep and narrow. This is known to result in a well-distributed generation of ZFs.

The basis of improved transport modelling lays in the field geometry. Other than using the residual to understand nonlinear ZF activity, the relevant dynamics in our observations happen during the early times of the linear ZF response. This can be analytically related to the geometric components of the field geometry. In short, our findings suggest a direct connection between ZF activity and certain key characteristic scales of the field geometry. The successful modelling of transport, we argue, depends on determining the quantitative relationship between these scales and nonlinear ZF activity, which will be the subject of future work.

The GENE simulations were performed on the MPCDF clusters (Germany) and the MARCONI supercomputer (Italy). This work has been carried out within the framework of the EUROfusion Consortium and has received funding from the Euratom research and training programme 2014-2018 and 2019-2020 under grant agreement No 633053. The views and opinions expressed herein do not necessarily reflect those of the European Commission.

\bibliographystyle{jpp}

\bibliography{Misc/library}

\end{document}